\documentclass{elsart}

\usepackage{graphicx}

\usepackage{amssymb}

\usepackage{enumerate}

\begin{document}

\begin{frontmatter}



\title{Modeling for evolving biological networks with scale-free connectivity, hierarchical modularity, and disassortativity}

\author[KIT,now]{Kazuhiro Takemoto},
\author[KIT,KIT2,cor1]{Chikoo Oosawa}
\ead{chikoo@bio.kyutech.ac.jp}

\corauth[now]{Current affiliation: Bioinformatics Center, Institute for Chemical Research, Kyoto University, Gokasho, Uji, Kyoto 611-0011, Japan.}
\corauth[cor1]{Corresponding author.}

\address[KIT]{Department of Bioscience and Bioinformatics, Kyushu Institute of Technology, Iizuka, Fukuoka 820-8502, Japan}
\address[KIT2]{Bioalgorithm Project, Faculty of Computer Science and Systems Engineering, Kyushu Institute of Technology, Iizuka, Fukuoka 820-8502, Japan}

\begin{abstract}
We propose a growing network model that consists of two tunable mechanisms: growth by merging modules which are represented as complete graphs and a fitness-driven preferential attachment. Our model exhibits the three prominent statistical properties are widely shared in real biological networks, for example gene regulatory, protein-protein interaction, and metabolic networks. They retain three power law relationships, such as the power laws of degree distribution, clustering spectrum, and degree-degree correlation corresponding to scale-free connectivity, hierarchical modularity, and disassortativity, respectively. After making comparisons of these properties between model networks and biological networks, we confirmed that our model has inference potential for evolutionary processes of biological networks.
\end{abstract}

\end{frontmatter}

\section{Introduction}
\label{sec:intro}
Structural data of large-scale biological networks such as gene regulatory networks \cite{Lee2002,Farkas2003,Luscombe2004}, protein-protein interaction networks \cite{Uetz2000,Ito2001,Giot2003,Raul2005}, and metabolic networks \cite{Jeong2000,Ma2002} have been accumulated by the active investigations in recent years, and heterogeneous connectivity, which means that a few nodes integrate a great number of nodes and most of the remaining nodes do not, are ubiquitously found as a striking property of the structures \cite{Farkas2003,Giot2003,Raul2005,Jeong2000,Ma2002,Albert2002,Barabasi2004,Albert2005,Colizza2005,Goh2005}.
Heterogeneity exhibits the scale-free feature \cite{Barabasi1999-1,Barabasi1999-2}, and differs from homogenous connectivity, which means that most nodes have the same degree (number of edges), is observed in classical network models such as regular lattices and the Erd\H{o}s-R\'enyi random network \cite{Erdos1960}. The scale-free feature is in degree distribution, defined as the probability of the existence nodes with degree $k$, which follows the power law, $P(k)\propto k^{-\gamma}$ with $2<\gamma<3$. Since the scale-free feature is independent of living species \cite{Jeong2000,Barabasi2004,Albert2005,EvoNet2003}, growth mechanism which can generate such statistical properties are paid much attention, and are expected to elucidate the evolutionary and growing processes of the networks.

In previous works, the Barab\'asi-Albert (BA) model \cite{Barabasi1999-2} is well-known as a model for scale-free networks, and is frequently embedded among variety of mathematical models \cite{Handbook2004}. Using the growth mechanism by adding new nodes with preferential attachment (PA) which means that high-degree nodes get an even better chance to attract new edges, the BA model generates scale-free networks which indicate power-law degree distributions with the fixed degree exponent $-3$; thus $P(k)\propto k^{-3}$ \cite{Barabasi1999-2,EvoNet2003,EvoInternet2004}. However, the recent detailed statistical analyses of biological networks manifest some discrepancies between the networks generated by the BA model and biological networks. In particular, there are two striking properties as follows.

One of the properties is a small-world feature \cite{Watts1998}. This feature is reflected in high clustering coefficients $C$, which denote the density of edges between neighbors of a given node, and implies modularity of networks \cite{Hartwello1999,Mering2003}. The modular structures are actively investigated with statistical approaches, and it is found that clustering spectra, defined as correlations between degree $k$ of a given node and the clustering coefficient $C$ of the node, follow the power law with exponent $-1$; thus $C(k)\propto k^{-1}$ \cite{Colizza2005,Goh2005,Ravasz2002}.
The power-law spectrum, observed in biological networks, suggests a hierarchical structure of modules \cite{Ravasz2002}. However, BA networks do not exhibit a power-law clustering spectrum, indicating the absence of the hierarchical structures \cite{Ravasz2003,Barrat2005}. In order to fill the gap, we propose a model (hereinafter called MM model) with growth mechanism by merging modules and PA from the BA model. The MM model shows power-law clustering spectra and power-law degree distributions with arbitrary degree exponents, demonstrating that our model can reproduce finer details of biological networks \cite{Takemoto2005}.  

The other of the properties is a heterogeneous degree-degree correlation \cite{Satorras2001-1}, defined as correlation between degree $k$ of a given node and average degree $\bar{k}_{nn}$ of neighbors of the node. In biological networks, the degree-degree correlations follow power law; $\bar{k}_{nn}(k)\propto k^{\nu}$ with a negative exponent, $-1<\nu<0$, empirically found \cite{Colizza2005,Goh2005,Yook2005}. In general, the properties are called disassortatvity \cite{Newman2002}. It is reported that the BA and MM model have no disassortativity \cite{Takemoto2005,Satorras2001-1}.
Pastor-Satorras {\em et al.} \cite{Satorras2001-1} and Barrat {\em et al}. \cite{Barrat2004-2} respectively argue that disassortativity emerges with competitive dynamics \cite{Bianconi2001} and weight-driven dynamics \cite{Barrat2004-1}, suggesting that growing mechanism needs to include fitness such as spatial distance \cite{Havlin2002}, aging \cite{Zhu2003}, and so on. 

In this paper, we propose a model which maintains the following statistical properties: the scale-free feature, the hierarchical modularity, and the disassortativity. For the proposal, we start with a modification of the MM model that has the scale-free feature and the hierarchical modularity \cite{Takemoto2005}. We embed a tunable fitness-driven (FD) mechanism into the MM model, and provide the numerical and analytical solutions for the statistical properties of model networks.  

\section{Model}
\label{sec:model}
Here we present an evolving network model. Our model includes growth by merging modules and the FD mechanism [see Section \ref{sec:disc} for biological implications of the model]. 

\begin{figure}[t]
\begin{center}
	\includegraphics{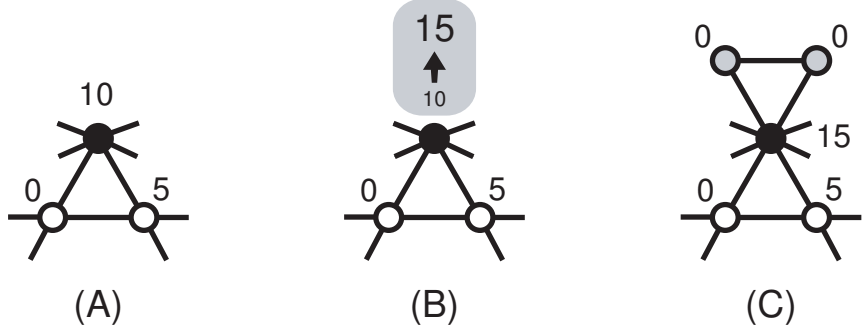}  
	\caption{Schematic diagram of growth process of our model network with $a=3$, $m=1$, and $\xi=5$. (A) Selection of node(s) by the FD-PA mechanism, Eq. (\ref{eq:PAf}). The filled node is selected by the attachment. Each number in the figure indicates the fitness of a corresponding node. (B) Updating of the fitness. The selected node's fitness increases according to the updating rule, Eq. (\ref{eq:fit-rule}). (C) Merging new modules. As a result, $(a-m)$ new node(s) are added, filled with gray, and take zero as their initial fitness.}  
	\label{fig:model}
\end{center}
\end{figure}

{\em Growth by merging modules} --- A network grows through joining new modules to existing nodes of a network over time [see Fig. \ref{fig:model} (C)]. $m/a$ is the first control parameter of the model where $a$ and $m$ denote the number of nodes of the complete graph and the number of merged node(s), respectively. Please note that the process develops without adding extra edges \cite{Takemoto2005}.

{\em Fitness-driven (FD) preferential attachment (PA)} --- The standard PA mechanism of the BA model is the probability that node $i$ is chosen to get an edge, and is proportional to the degree of node $i$; hence $\Pi^{\mathrm{BA}}_i=k_i/\sum_jk_j$, where $k_i$ is the degree of node $i$. The mechanism only considers the degrees at the nodes. Here, we additionally consider the probability that node $i$ is selected according to degree $k_i$ and fitness $f_i$, and express the probability as
\begin{equation}
\Pi_i=\frac{k_i+f_i}{\sum_j(k_j+f_j)}.
\label{eq:PAf}
\end{equation}

{\em Updating rule of fitness} --- Moreover, we consider the change of the fitness of node $i$. When node $i$ is selected by using the FD-PA mechanism given in Eq. (\ref{eq:PAf}), the fitness of node $i$ increases as follows:
\begin{equation}
f_i\gets f_i+\xi,
\label{eq:fit-rule}
\end{equation}
where $\xi$ takes a constant value, and is the second control parameter in the model, indicating the strength of incidence of the fitness.

Taken together, our model networks are generated by the following procedures.
\begin{enumerate}[i)]
\item We start from a module that is a complete graph consisting of $a$ ($\geq 3$) nodes. For fitness, we assign zero to all nodes in the module.
\item At every time step a new module with the same size is joined by merging to an existing $m$ $(<a)$ nodes.
\item When merging the module, the FD-PA mechanism, Eq (\ref{eq:PAf}), is used to select $m$ old nodes [see Fig. \ref{fig:model} (A)]. Then, the fitnesses of the old nodes are increased using the updating rule, Eq. (\ref{eq:fit-rule}) [see Fig. \ref{fig:model} (B)]. At the end, zero is assigned to the fitnesses of the new nodes [see Fig. \ref{fig:model} (C)]. Resultant duplicated edges between the merged nodes are counted and contribute to the FD-PA mechanism in the succeeding steps.
\end{enumerate}

When $a>m$ and $\xi\geq 0$, the model network evolves in time steps. In addition, our model is equivalent to the BA model with the specific condition: $a=2$, $m=1$, and $\xi=0$.

\section{Degree distribution}
\label{sec:degree}
In this section, we present the analytical and the numerical solutions for the degree distribution of our model. The degree distribution is an important statistical property for the characterization of the networks, and denotes the existence probability of nodes with degree $k$. The degree distribution \cite{EvoNet2003,EvoInternet2004} is defined as
\begin{equation}
P(k)=\frac{1}{N}\sum_{i=1}^N\delta(k_i-k),
\label{eq:P(k)_def}
\end{equation}
where $\delta(x)$ and $N$ are Kronecker's delta function and the total number of the nodes, respectively.

\subsection{Analytical solution}
\label{subsec:deg-ana}
In order to describe the degree distribution, we take the continuous mean-field approach \cite{Barabasi1999-2,Barrat2005}. The standard approach can not be applied directly due to the inclusion of the fitness updating. We express the degree and the fitness as  
\begin{equation}
F_i=k_i+f_i,
\label{eq:F}
\end{equation}
and investigate a time evolution of $F_i$. Since the fitness updating rule is represented as $f_i\gets f_i+\xi$, $F_i$ satisfies
\begin{equation}
F_i=\left(\frac{\xi}{a-1}+1\right)k_i-\xi,
\label{eq:F-k}
\end{equation} 
indicating the proportional relationship between $F_i$ and $k_i$.

The time evolution of $F_i$ is described as
\begin{equation}
\frac{dF_i}{dt}=m(a-1+\xi)\frac{F_i}{\sum_jF_j},
\label{eq:dF/dt}
\end{equation}
where $\sum_jF_j\approx [a(a-1)+m\xi]t$. The solution of the equation with $F_i(t=s)={a \choose 2}+\xi=A(a,\xi)$ as an initial condition for Eq. (\ref{eq:dF/dt}) is
\begin{equation}
F_i(t)=A(a,\xi)\left(\frac{t}{s}\right)^\beta,
\label{eq:F(t)}
\end{equation}
where $\beta=[m(a-1+\xi)]/[a(a-1)+m\xi]$. Because $s/t$ denotes the probability that $F_i$ is larger than a given $F$, Equation (\ref{eq:F(t)}) is rewritten as
\begin{equation}
P(>F)=A(a,\xi)^{1/\beta}F^{-1/\beta},
\label{eq:preP(F)}
\end{equation} 
which corresponds to the cumulative probability. From Eq. (\ref{eq:preP(F)}), the probability distribution for $F$ is given as
\begin{equation}
P(F)=-\frac{d}{dF}P(>F)=\frac{A(a,\xi)^{1/\beta}}{\beta}F^{-\gamma},
\label{eq:P(F)}
\end{equation}
where $\gamma=(1/\beta)+1$. Finally, substituting Eq. (\ref{eq:F-k}) into Eq. (\ref{eq:P(F)}), we get the degree distrubution
\begin{equation}
P(k)\simeq B(a,\xi,\gamma)k^{-\gamma},
\label{eq:P(k)}
\end{equation}
where $B(a,\xi,\gamma)=A(a,\xi)^{\gamma-1}[\xi/(a-1)+1]^{-\gamma}/(\gamma-1)$. The degree distribution obeys the power-law with the degree exponent
\begin{equation}
\gamma=\frac{(a+m)(a-1)+2m\xi}{m(a-1+\xi)},
\label{eq:gamma}
\end{equation}
demonstrating that our model network represents the scale-free feature.

\subsection{Numerical solution}
\label{subsec:deg-num}
\begin{figure}[t]
\begin{center}
	\includegraphics{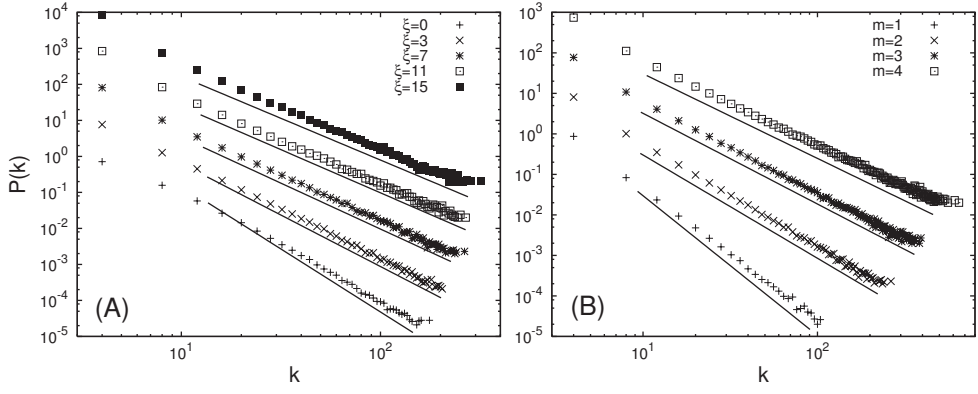}  
	\caption{Degree distributions $P(k)$ with $a=5$ and $N=50000$ (shifted for clarity). Different symbols correspond to the different numerical results. Solid lines show relationship $\propto k^{-\gamma}$, where $\gamma$ is predicted by Eq. (\ref{eq:gamma}). (A) $\xi$ dependency with $m=2$. (B) $m/a$ dependency with $\xi=7$.}
	\label{fig:degree}
\end{center}
\end{figure}

In order to confirm the analytical predictions, we performed numerical simulations of networks generated by using our model described in Sec. \ref{sec:model}. We show degree distributions with $a=5$, $N=50000$, different $m\in[1,4]$ and $\xi\in[0,15]$ in Fig. \ref{fig:degree}. The degree distributions follow the power law, reflecting the scale-free feature.   

Figure \ref{fig:degree} (A) shows the degree distributions with $m=2$ and the different $\xi$. The degree exponents decay with increasing $\xi$. Figure \ref{fig:degree} (B) shows the degree distributions with $\xi=7$ and the different $m$. The degree exponents decay with increasing $m$. We show excellent agreement between the numerical results and the theoretical predictions, demonstrating that $\gamma$ is function of $m/a$ and $\xi$.

\section{Degree entropy}
As shown in Eq. (\ref{eq:gamma}) and Fig. \ref{fig:degree}, the degree exponent $\gamma$ is variant with $\xi$ and $m/a$. We consider the degree entropy \cite{Sole2004} to characterize parameter dependency for the heterogeneous connectivity of our model. Assuming that $\sum_{k=1}^{N-1}P(k)=1$, the degree entropy is defined as
\begin{equation}
H=-\sum_{k=1}^{N-1}P(k)\ln P(k),
\label{eq:H}
\end{equation}
and provides an average measure of a network's heterogeneity because it relies on the diversity of the degree distribution. The increment of degree entropy $H$ corresponds to the increment of the heterogeneity for complex networks \cite{Sole2004}.

\begin{figure}[t]
\begin{center}
	\includegraphics{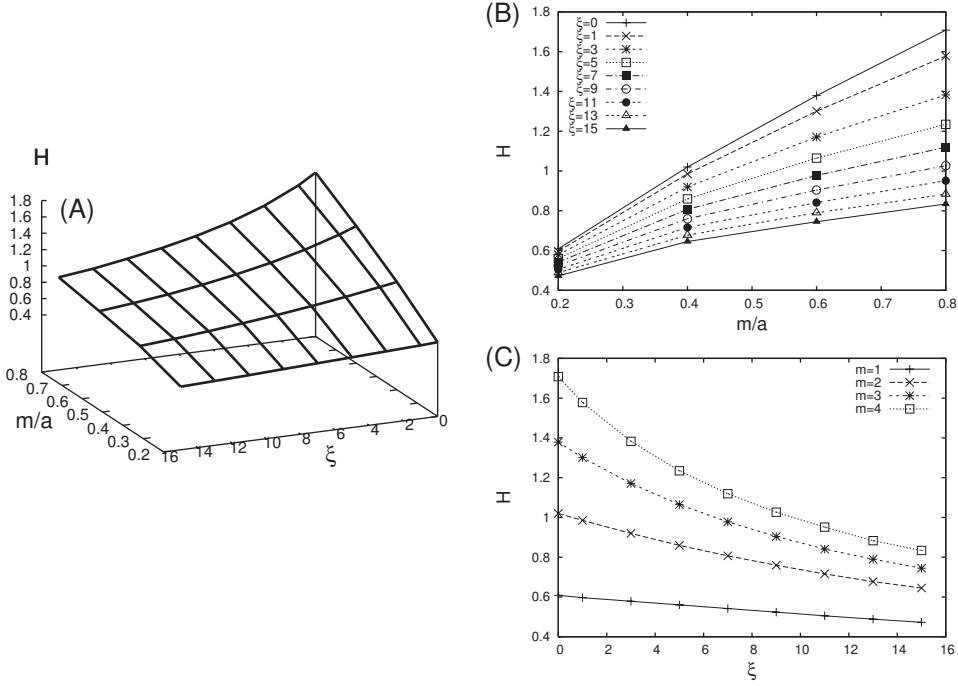}  
	\caption{Degree entropy $H$ with $a=5$ and $N=12800$ (Axis direction changed for clarity). (A) 3D plot of the entropy versus the variables $\xi$ and $m/a$. Projections of the plot to the $(m/a)$-$H$ plane (B) and the $\xi$-$H$ plane (C).}
	\label{fig:H}
\end{center}
\end{figure}

Figure \ref{fig:H} shows the numerical results of the degree entropy of the model network with $a=5$, $N=12800$, different $m\in[1,4]$, and different $\xi\in[0,15]$.

In Fig \ref{fig:H} (A), we show the degree entropy $H$ with the different $m/a$ and $\xi$ in three dimensions. The degree entropy $H$ increases for the larger $m/a$ and the smaller $\xi$, and decreases in the opposite direction. In Fig \ref{fig:H} (B), the projections of the plot to the $(m/a)$-$H$ plane of Fig. \ref{fig:H} (A) are shown. 
The correlation between $H$ and $m/a$ becomes strong with increasing $\xi$. In Fig \ref{fig:H} (C), the projections of the plot to the $\xi$-$H$ plane of Fig. \ref{fig:H} (A) are shown. The degree entropy decays with increasing $\xi$ in the case of constant $m/a$.

\section{Clustering spectrum}
\label{sec:C(k)}
The clustering spectrum is a well-known statistical property which reflects the hierarchical modularity of networks \cite{EvoNet2003,EvoInternet2004}, and is defined as
\begin{equation}
C(k)=\frac{\sum_{i=1}^NC_i\times\delta(k-k_i)}{NP(k)},
\label{eq:C(k)}
\end{equation}
where $\delta(x)$ denotes Kronecker's delta function, and $C_i$ corresponds to the clustering coefficient \cite{Watts1998}, defined as
\begin{equation}
C_i=\frac{M_i}{{k_i\choose 2}}=\frac{2M_i}{k_i(k_i-1)},
\label{eq:C}
\end{equation}
and represents the density of edges among neighbors of a given node, where $k_i$ and $M_i$ denote the degree of node $i$ and the number of edges among the neighbors, respectively.

First, we give the analytical solution for the clustering spectrum of our model. Since our model grows through merging modules (see Sec. \ref{sec:model} and Fig. \ref{fig:model}), the number of edges of node $i$ among neighboring nodes is approximately described as \cite{Takemoto2005}
\begin{equation}
M_i\simeq S_i{a-1\choose 2}=S_i(a-1)\frac{a-2}{2},
\label{eq:MM_M}
\end{equation}
where $S_i$ corresponds to the number of the selection of node $i$ with the PA,  as in Eq (\ref{eq:PAf}). In our model, because the degree of node $i$ is expressed as $k_i=S_i(a-1)$, Eq. (\ref{eq:MM_M}) is rewritten as
\begin{equation}
M_i\simeq \frac{a-2}{2}k_i,
\label{eq:MM_M2}
\end{equation}
indicating the proportional relationship between $M_i$ and $k_i$. Finally, substituting Eq. (\ref{eq:MM_M2}) into Eq. (\ref{eq:C}), we get the clustering spectrum
\begin{equation}
C(k)\simeq \frac{a-2}{k} \propto k^{-1}.
\label{eq:C(k)_MM}
\end{equation}
The clustering spectrum follows the power law with the exponent $-1$ when $a\geq 3$, reflecting the hierarchical modularity of our model network.

\begin{figure}[t]
\begin{center}
	\includegraphics{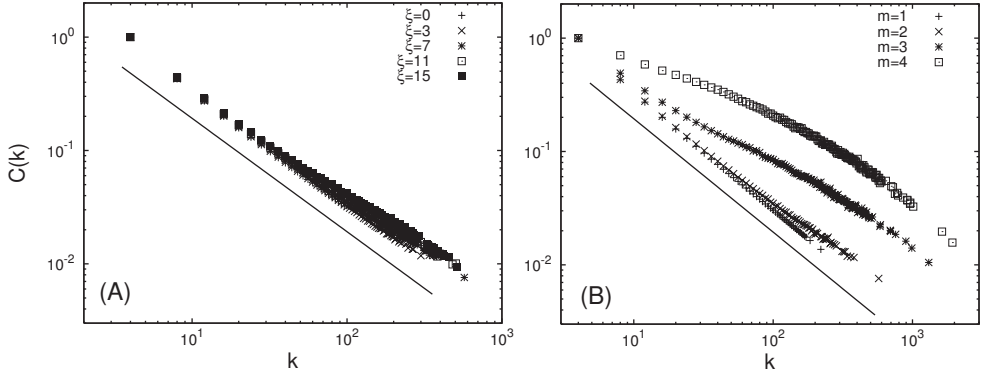}  
	\caption{Cluster spectra $C(k)$ with $a=5$ and $N=50000$. Different symbols correspond to the numerical results. Solid lines show the relationship $\propto k^{-1}$. (A) $\xi$ dependency with $m=2$. (B) $m$ dependency with $\xi=7$.}
	\label{fig:C(k)}
\end{center}
\end{figure}

Next, we present the numerical results of our model for the clustering spectrum to verify the analytical solution. In Fig. \ref{fig:C(k)}, we show clustering spectra with $a=5$, $N=50000$, different $m\in[1,4]$, and different $\xi\in[0,15]$.

Figure \ref{fig:C(k)} (A) shows the clustering spectra with $m=2$ and the different $\xi$. The spectra follow power law with the exponent $-1$ despite changing $\xi$. Figure \ref{fig:C(k)} (B) shows the clustering spectra with $\xi=7$ and the different $m$. The spectra follows power law with the exponent $-1$ for large $k$. For small degree $k$, the cut-off is prominent with increasing $m$ due to running away from the approximation of Eq. (\ref{eq:MM_M}).

The clustering spectra follow the power law with the exponent more or less equal to $-1$, reflecting the hierarchical modularity of networks.

\section{Global clustering coefficient}
\label{sec:C(N)}
As seen in Fig. \ref{fig:H}, degree entropy $H$ is sensitive to $\xi$ and $m/a$. To examine parameter-dependency of the modularity (clustering property) for the whole network, we utilize a global clustering coefficient $C$. The coefficient is defined as \cite{Watts1998}
\begin{equation}
C=\frac{1}{N}\sum_{i=1}^NC_i.
\label{eq:GC}
\end{equation}
A higher value of $C$ means that the network has higher modularity. 

\begin{figure}[t]
\begin{center}
	\includegraphics{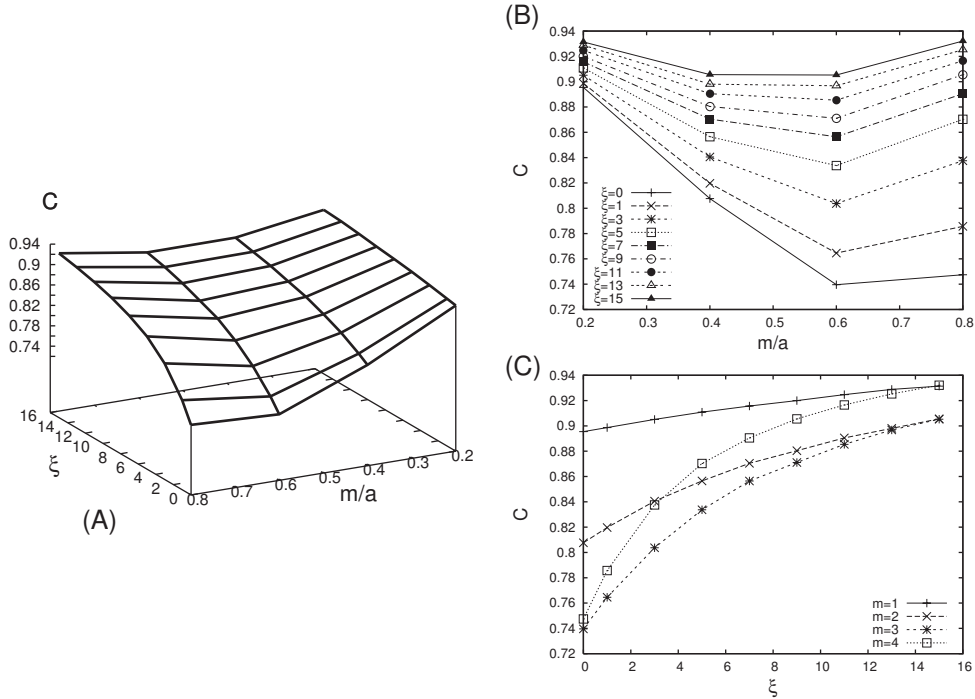}  
	\caption{Global clustering coefficient $C$ with $a=5$ and $N=12800$. (A) 3D plot of the coefficient versus the variables, $\xi$ and $m/a$ (axis direction changed for clarity). Projections of the plot to the $(m/a)$-$C$ plane (B) and the $\xi$-$C$ plane (C).}
	\label{fig:C}
\end{center}
\end{figure}

Figure \ref{fig:C} shows the numerical result of $C$ in our model network with $a=5$, $N=12800$, different $m\in[1,4]$, and different $\xi\in[0,15]$. 

In Fig. \ref{fig:C} (A), we show the $C$ with different $m/a$ and $\xi$ in three dimensions. $C$ tends to increase more for smaller $m/a$ and larger $\xi$.

Since the small or large $m/a$ give rise to different effects on $C$, Figure \ref{fig:C} (A) has a valley in the middle of $m/a$. In the case of small $m/a$, $C$ can remain high because the modules combine via few common nodes. In the case of large $m/a$, in contrast, $C$ leniently decreases because the networks tend to be randomized since the modules combine via many common nodes. In this case, however, FD mechanism helps to increase the fitness of the nodes in the modules, inducing a formation of a cluster with high-edge density. As a result, $C$ increases with $m/a$. Due to the trade-off mechanism, the biphasic graph is shown in Fig. \ref{fig:C} (B). In addition, $C$ is less sensitive for large $\xi$. In Fig. \ref{fig:C} (C), we show the projection of the plot to the $\xi$-$C$ plane of Fig. \ref{fig:C} (A).  

\section{Degree-degree correlation}
\label{sec:knn(k)}
The degree-degree correlation is a statistical property that characterizes assortativity of the networks, and represents the average degree of neighbors of nodes with degree $k$ \cite{EvoNet2003,EvoInternet2004}. The correlation is defined as 
\begin{equation}
\bar{k}_{nn}(k)=\sum_{k'}k'P(k'|k),
\label{eq:knn(k)}
\end{equation}
where the conditional probability $P(k'|k)$ is the frequency that a node with degree $k$ connects to a node with degree $k'$. Using Kronecker's delta function, we redefined the degree-degree correlation as
\begin{equation}
\bar{k}_{nn}(k)=\frac{\sum_{i=1}^{N}\langle k_{nn}\rangle_i\times \delta(k_i-k)}{NP(k)},
\label{eq:knn(k)-2}
\end{equation}
where $\langle k_{nn}\rangle_i$ denotes the average nearest-neighbor degree, written as
\begin{equation}
\langle k_{nn}\rangle_i=\frac{1}{k_i}\sum_{h\in V(i)}k_h,
\label{eq:knn}
\end{equation}
where $V(i)$ corresponds to the set of neighbors of node $i$.

\begin{figure}[t]
\begin{center}
	\includegraphics{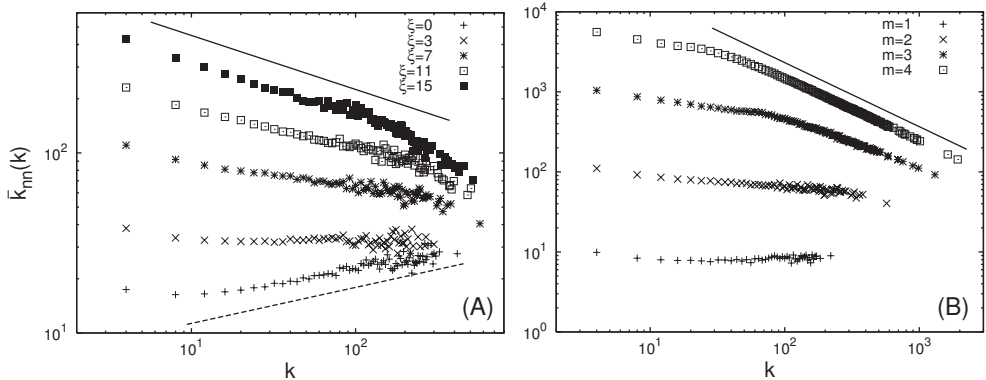}  
	\caption{Degree correlations $\bar{k}_{nn}(k)$ with $a=5$ and $N=50000$. (A) $\xi$ dependency with $m=2$. A solid and dashed line correspond to $\propto k^{-0.3}$ and $\propto k^{0.2}$, respectively. (B) $m$ dependency with $\xi=7$. A solid line shows the relationship $\propto k^{-0.8}$.}
	\label{fig:corr}
\end{center}
\end{figure}

Here, we give the numerical solutions for the degree-degree correlation of our model. Figure \ref{fig:corr} shows the degree-degree correlations with $a=5$, $N=50000$, the different $m$, and $\xi$. The correlations follow the power law; this $\bar{k}_{nn}(k)\propto k^{\nu}$ with $-1<\nu<0$ as a roughly observation. Negative and larger $\nu$ tend to be seen for the larger $m/a$ and $\xi$. 

Figure \ref{fig:corr} (A) shows the degree-degree correlations with $m=2$ and the different $\xi$. The exponent $\nu$ decays with increasing $\xi$. Figure \ref{fig:corr} (A) shows the degree-degree correlations with $\xi=7$ and the different $m$. The exponent $\nu$ decays with increasing $m$.

Due to the fitness updating and the PA, the degree-degree correlations follow the power law, reflecting the disassortativity of the networks. As previously reported, the disassortativity is not reproduced if we consider the PA mechanism only \cite{Takemoto2005,Satorras2001-1}.

\section{Assortative coefficient}
\label{sec:r}
The assortative coefficient (AC) \cite{Newman2002} is can be thought of as a compendium parameter of the degree-degree correlations, and is defined as
\begin{equation}
r=\frac{4\langle k_ik_j\rangle-\langle k_i+k_j\rangle^2}{2\langle k_i^2+k_j^2\rangle-\langle k_i+k_j\rangle^2}
\label{eq:r}
\end{equation}
where $k_i$ and $k_j$ are the degrees of two vertices at the ends of an edge, and $\langle \cdots \rangle$ denotes the average over all edges. In other words, the AC is the correlation coefficient for the degree-degree correlation $\bar{k}_{nn}(k)$, and takes values $-1\leq r\leq 1$. The relationship between the AC and network structures is described as follows:

\begin{enumerate}[i)]
\item In the case of $r<0$, low-degree nodes tend to connect to high-degree nodes, indicating the disassortativity. Then, the degree-degree correlation $\bar{k}_{nn}(k)$ decreases with increasing $k$. 

\item In the case of $r=0$, the degree-degree correlation $\bar{k}_{nn}(k)$ is uncorrelated.

\item In the case of $r>0$, high-degree nodes tend to connect to high-degree nodes, reflecting the assortativity. Then, the degree-degree correlation $\bar{k}_{nn}(k)$ increases with degree $k$.
\end{enumerate}

\begin{figure}[t]
\begin{center}
	\includegraphics{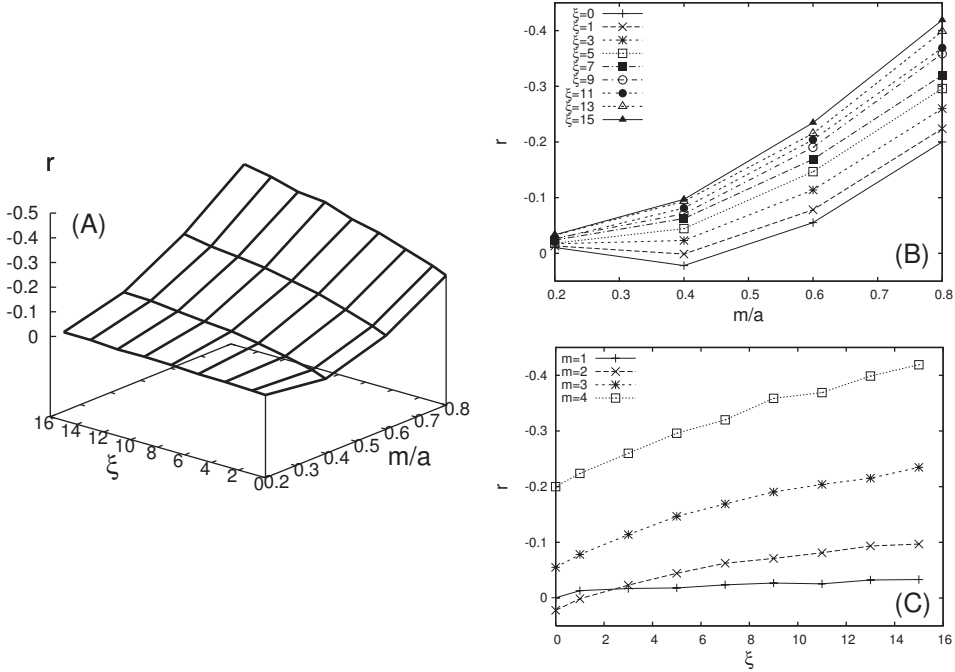}  
	\caption{Assortative coefficient $r$ with $a=5$ and $N=12800$. (A) 3D plot of the coefficient versus the variables, $\xi$ and $m/a$. Projections of the plot to the $(m/a)$-$r$ plane (B) and the $\xi$-$r$ plane (C). Vertical axis of (A)--(C) is inverted for clarity.}
	\label{fig:r}
\end{center}
\end{figure}

Figure \ref{fig:r} shows the numerical solutions of the AC for our model with $a=5$, $N=12800$, different $m\in[1,4]$, and $\xi\in[0,15]$.

Figure \ref{fig:r} (A) shows the AC with the different $m/a$ and $\xi$ in three dimensions. The large negative AC generally tends to be seen for the larger $m/a$ and $\xi$. Figure \ref{fig:r} (B) shows the projections of the plot to the $(m/a)$-$r$ plane of Fig. \ref{fig:r} (A). Two biphasic curves (once $r$ takes positive instead of negative) for the smaller $\xi\in[0,1]$ emerge because positive degree-degree correlations are exhibited in the case of $m/a \leq 0.5$ \cite{Takemoto2005}. For larger $\xi\in[3,15]$, assortative coefficients $r$ monotonically decrease (upward) with increasing $m/a$. Figure \ref{fig:r} (C) shows the projections of the plot on the $\xi$-$r$ plane for Fig. \ref{fig:r} (A).

\section{Comparison of statistical properties}
\label{sec:compari}
In order to validate our model we make a comparison of statistical properties between biological networks and our model networks. We prepare two different types of networks; the gene regulatory \cite{Salgado2004} and the metabolic \cite{Jeong2000} networks of {\em Escherichia coli}. The gene regulatory network is represented as sets of graphs consisting of nodes and edges; they correspond to genes and interactions among genes, respectively. For simplicity, we extract the largest component from the networks, and replace directed and/or weighted edges with undirected and unweighted edges. Moreover, we remove multiple edges and self-loops. The metabolic network is transformed with the same procedures.

\begin{figure}[t]
\begin{center}
	\includegraphics{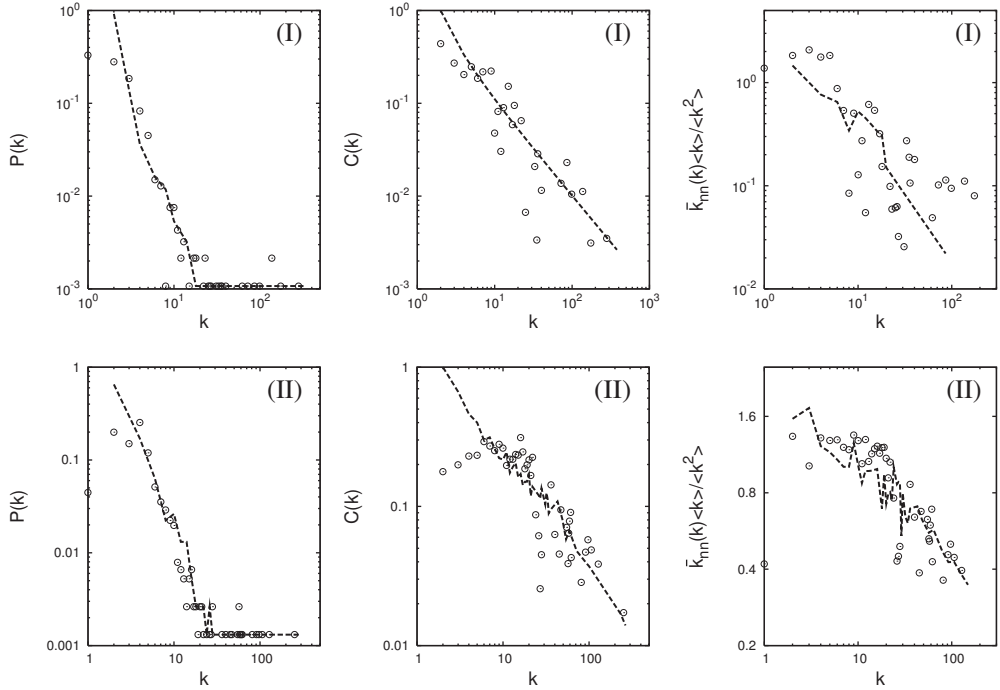}  
	\caption{Comparison of statistical properties between biological networks and model networks. The left column shows degree distributions $P(k)$. Clustering spectra $C(k)$ are in center column. Degree-degree correlations $\bar{k}_{nn}(k)$ constitute the right column. Note that the degree-degree correlations $\bar{k}_{nn}(k)$ are divided by $\langle k^2\rangle/\langle k \rangle$ for normalization. Symbols and dashed lines denote data of biological networks and our model networks, respectively. (I) Gene regulatory network in {\it Escherichia coli} \cite{Salgado2004} and our model with $a=3$, $m=1$, and $\xi=20$. (II) Metabolic networks in {\it Escherichia coli} \cite{Jeong2000} and our model with $a=3$, $m=2$, and $\xi=1$. The size of the model network is the same as total number of nodes of the biological network.}
	\label{fig:compari}
\end{center}
\end{figure}

In Fig. \ref{fig:compari}, we show the statistical properties of the biological networks and our model networks. We have good agreement with the real network's data and our model, demonstrating that our model reproduces the three striking statistical properties, also widely shared in the biological networks. In addition we confirm similar relationships with network data derived from the yeast \cite{Jeong2001} and the worm \cite{Li2004} (data not shown).

\section{Discussions}
\label{sec:disc}
Even though large amounts of structural data of biological networks have been available in recent years, little is known about the detailed mechanisms behind the origin and evolution of the networks. 
However, we proposed a hypothetical model and its parameter space that can reproduce biological features from a statistical viewpoint. We think that the growth mechanism by merging modules and the FD-PA mechanism are important for evolutionary processes in biological networks.

Since new genes commonly appear through complete or partial duplication of pre-existing genes \cite{Rzhetsky2001,Alves2002,Eisenberg2003,Pastor2003,Teichmann2004}, the resultant sequences are subject to selective pressure as parts of the whole system and would give rise to changes in functional interactions among biomolecules, resulting in changes of statistical properties of network topology.

In the case of metabolic networks, modules in our model can be regarded as a set of chemical compounds (nodes) and pathways (edges), and the merging process may correspond to reminiscent tinkering processes such as bridging pathways between pre-existing nodes, enabling doubling pathways. These processes may positively or negatively contribute to increase robustness against mutations or give divergence and become a new pathway. The assumption of modular elements in our model is supported because module structures are found in many biological networks \cite{Mering2003,Wuchty2003}.

The original PA mechanism is exactly proportional to the number of edges of the node \cite{Barabasi1999-2}. The FD-PA mechanism can consider additional or extra fitness of nodes. Rozenfeld {\it et al.} \cite{Havlin2002} and Zhu {\it et al.} \cite{Zhu2003} introduce into nodes the additional factors aging and a radius from a node in two-dimensional space. Because the additional factors are determined on the basis of a node's property, it is easy for us to infer that the effects from the factor is considered to be underlie physicochemical properties of biomolecules such as mass, length, affinity, structural stabilities, and contents of bases or amino acids in a biomolecule. Together with the idea increasing fitness during the growth process reflects a certain variation of node's properties with time, and is relevant to specification and/or fixation process after duplication and divergence.

\section{Summary}
\label{sec:conc}
We have presented analytical and numerical results verifying that our model produces the three remarkable statistical properties that are widely shared in biological network structures.

{\em Scale-free connectivity} --- The degree distribution of our model follows power law, indicating the scale-free property; thus $P(k)\propto k^{-\gamma}$. The degree exponent $\gamma$ can be predicted by Eq. (\ref{eq:gamma}), and demonstrates the range $2<\gamma<\infty$ which includes the range of the degree exponents that are observed in biological networks $2<\gamma<3$ \cite{Albert2002,Albert2005,EvoNet2003}. The degree exponents take smaller values, meaning that the degree distribution becomes flat for the larger $m/a$ and $\xi$.

{\em Hierarchical modularity} --- The clustering spectrum of our model follows the power law with the exponent $-1$; hence $C(k)\propto k^{-1}$, reflecting the hierarchical modularity \cite{Ravasz2002,Ravasz2003}. The power-law spectrum is nearly independent of $\xi$ and $m/a$. The modularity of the whole network is higher for large $\xi$ and smaller $m/a$. The hierarchy manifests growing mechanism by merging modules.

{\em Disassortativity} --- The degree-degree correlation of our model follows the power law; therefore $\bar{k}_{nn}(k)\propto k^{\nu}$, reflecting the disassortativity. The exponent $\nu$ has the range $-1<\nu<0$ which includes the range of the exponents that are seen in biological networks \cite{Colizza2005,Goh2005,Yook2005}. The result indicates that the FD mechanism with positive $\xi$ is critical to reproduce this property. In addition, the disassortativity is sensitive to $\xi$ and $m/a$.



\section*{Acknowledgement}
This work was partially supported by Grant-in-Aid No.18740237 from MEXT (JAPAN).

\end{document}